# Neutron Emission Spectrometer To Measure Ion Temperature On The Fusion Demonstration Plant

P. J. F. Carle,[1,a] F. Retière,[2] A. Sher,[2] R. Underwood,[2] K. Starosta,[3] M. Hildebrand,[1] S. Barsky[1], S. Howard[1]

[1]*General Fusion Inc., 108 – 3680 Bonneville Place, Burnaby, British Columbia V3N 4T5, Canada*
[2]*TRIUMF, 4004 Wesbrook Mall, Vancouver, British Columbia V6T 2A3, Canada*
[3]*Department of Chemistry, Simon Fraser University, 8888 University Dr, Burnaby, British Columbia V5A 1S6, Canada*

[a]Author to whom correspondence should be addressed: pat.carle@generalfusion.com.

General Fusion is building the Fusion Demonstration Plant to demonstrate a magnetized target fusion scheme in which a deuterium plasma is heated from 200 eV to 10 keV by piston-driven compression of a liquid-lithium liner. The multilayer coaxial time-of-flight (MCTOF) neutron emission spectrometer is designed to measure the ion temperature near peak compression at which time the neutron yield will approach $10^{18}$ neutrons/s. The neutron energy distribution is expected to be Gaussian since the machine uses no neutral beam or radio-frequency heating. In this case, analysis shows that as few as 500 coincidence events should be sufficient to accurately measure the ion temperature. This enables a fast time resolution of 10 μs, which is required to track the rapid change in temperature approaching peak compression. We overcome the challenges of neutron pile-up and event ambiguity with a compact design having two layers of segmented scintillators. The error in the ion temperature measurement is computed as a function of the neutron spectrometer's geometric parameters and used to optimize the design for the case of reaching 10 keV at peak compression.

## I. INTRODUCTION

General Fusion is working towards building the Fusion Demonstration Plant (FDP) in Culham, U.K. with operations scheduled to begin in 2025. The FDP will use a Magnetized Target Fusion (MTF) [1] scheme to compress a deuterium plasma to fusion conditions. A conceptual drawing is shown in Fig. 1.

Inside the main vacuum vessel, the machine's rotor spins up a liquid lithium liner to create a 3 m diameter cylindrical cavity. An array of pistons is fired with a relative delay between rows to initiate the spherical collapse of the liner.

The FDP will use a magnetized Marshall gun [2] at the top of the device to form a spherical tokamak plasma configuration via fast coaxial helicity injection [3] with the initial plasma having expected parameters of T ~ 200 - 400 eV and $n$ ~$5 \times 10^{19}$ m⁻³. Current running through a solid metal shaft along the geometric axis of the machine generates toroidal flux to control the q-profile of the plasma. The plasma is heated first by internal Ohmic decay and then reaches fusion conditions through rapid, near-adiabatic compressional heating with corresponding density increase.

The compression is timed to begin once the plasma has stabilized in the cavity. The plasma reaches peak compression in approximately 3-5 ms at which point the now spherically shaped liner cavity has a diameter of 30 cm. At peak compression, the plasma temperature and density are expected to increase to 10 keV and $3 \times 10^{22}$ m⁻³ respectively.

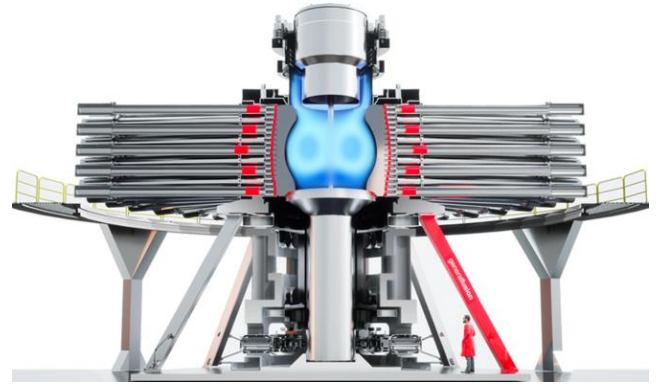

*Fig. 1. FDP conceptual drawing.*

The ion temperature, $T_i$, is a key parameter needed to evaluate the success of a compression. Three main diagnostic systems are being designed to measure $T_i$: ion Doppler spectroscopy, neutron yield, and neutron emission spectroscopy. Other $T_i$ diagnostics have been considered such as collective Thomson scattering and neutral beam charge exchange spectroscopy, but access issues are currently disqualifying them from further consideration.

Access to the plasma will be a challenge for many FDP diagnostics. Obstruction-free sightlines for conventional spectroscopic or laser-based diagnostics will only be possible via the top and bottom of the machine since the liquid lithium liner will block the view to the plasma from the sides. As compression proceeds, the liner will occlude an increasing number of sight lines. An example compression trajectory is given in Fig. 2.

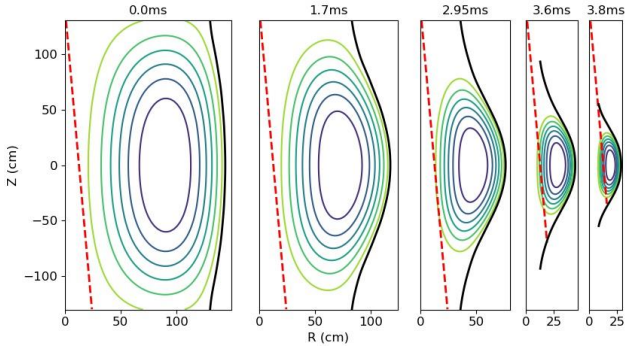

*Fig. 2. Example compression trajectory with liquid lithium liner and poloidal flux contours shown as solid lines. Neutron emission spectrometer line of sight is dashed. This shows a time sequence during the compression from left to right of t = 0, 1.7, 2.95, 3.6, and 3.8 ms, respectively.*

## II. Neutron Emission Spectroscopy

Fusion neutrons emerging from an MTF plasma will have a Gaussian energy distribution that is a function [4] [5] of the temperature of the reacting ions:

$$\sigma_{T_i} = \frac{82.5\sqrt{T_i}}{2\sqrt{2ln(2)}} \quad (1)$$

where $\sigma_{T_i}$ is the standard deviation of the neutron energy distribution in keV and $T_i$ is in keV. A neutron emission spectrometer (NES) measures the spread in the neutron energy distribution to estimate $T_i$.

Neutrons that have scattered before reaching the NES will contaminate the measurement with a low-energy population of detection events. A neutron collimator, consisting of a thick neutron absorbing material with a long central hole, is needed to shield the NES from this background of lower energy scattered neutrons; so what reaches the NES is predominantly a beam of directed neutrons sourced from the unobstructed line of sight passing through the plasma. Design of neutron collimators that sufficiently reduce the ambient background of scattered neutrons is well understood in the field of radiography [6] and for fusion diagnostic applications [7] [8].

In the case of a deuterium plasma, the collimated neutron beam has an average energy of 2.45 MeV with a thermal spread due to plasma ion temperature. An advantage of this MTF scheme is that the neutron energy spectrum should be a simple Gaussian since the FDP has no neutral beam or radio-frequency heating.

There are several possible NES techniques [9] [10], some of which are briefly highlighted below.

A compact detector (e.g. diamond) is a device in which an incident neutron deposits some of its energy and generates a pulse of some height and shape that can be measured. The properties of a given pulse depend on the incident neutron energy, the deposited energy, and type of interaction between the neutron and detector material. The pulse height is related to the incident neutron energy but, to be useful as a spectrometer, the detector's complex response must be characterized over the range of possible incident neutron energies and deposited energies.

In a magnetic proton recoil (MPR) neutron spectrometer, a hydrogen-rich thin foil is placed in the path of a collimated neutron beam originating from the plasma. Neutron collisions with the foil generate recoil protons that can be deflected by a magnetic field to an array of detectors to analyze their momentum.

A time-of-flight (TOF) neutron emission spectrometer [7] consists of a neutron collimator and two groups of scintillators, Layer 1 and Layer 2 as shown in Fig. 4. The beam of fusion neutrons is first incident on scintillator(s) in Layer 1, and neutrons either pass through the material undetected, or collide with a scintillator proton creating a burst of light that is detected. A deflected neutron heading toward the ring of detectors in Layer 2 has some chance of scattering again within Layer 2. The time between correlated scattering events in Layers 1 and 2 is related to incident neutron energy. After some integration time, a distribution of neutron energies will emerge.

In several previous TOF spectrometer designs, Layer 2 scintillators are positioned tangent to the surface of the sphere of constant TOF [7] [8]. When a neutron collides with a proton in Layer 1, the neutron will exit with an energy that depends on its angle of deflection. Large-angle deflections result in low exit-energies making it so that a trajectory in any direction will intersect the surface of this special sphere after a fixed period of time. The transit time to pass across the sphere of constant TOF will only depend on the incident energy of the neutron before the first scattering. Positioning moderately large detector plates tangent to this sphere will result in very small errors in the estimate of the original neutron energy because scintillation hits anywhere in that large detector plate are nearly equivalent to each other in terms of the transit time across the sphere. With this simplifying principle it is possible to construct a high-resolution spectrometer with only a modest number of detectors.

Diagnostic complications arise in a successful MTF compression scenario, where the ion temperature and neutron yield will rapidly increase many orders of magnitude during compression (Fig. 3). This poses a very different diagnostic challenge compared to the nearly steady state fusion rate in a tokamak. To understand the conditions achieved near peak compression, it is required to accumulate neutron energy spectra on a timescale of 10 μs and have good enough statistics in each spectrum to estimate $T_i$ with a 10% uncertainty or less. The expected high count rate at peak neutron yield requires a strategy that avoids pileup in any one detector, while making a choice of overall size that optimizes efficient coincidence detection. The proposed solution [11] discards the sphere of constant TOF concept in favor of many small, segmented scintillators in a compact arrangement to maximize the number of useful events while maintaining a high energy resolution.

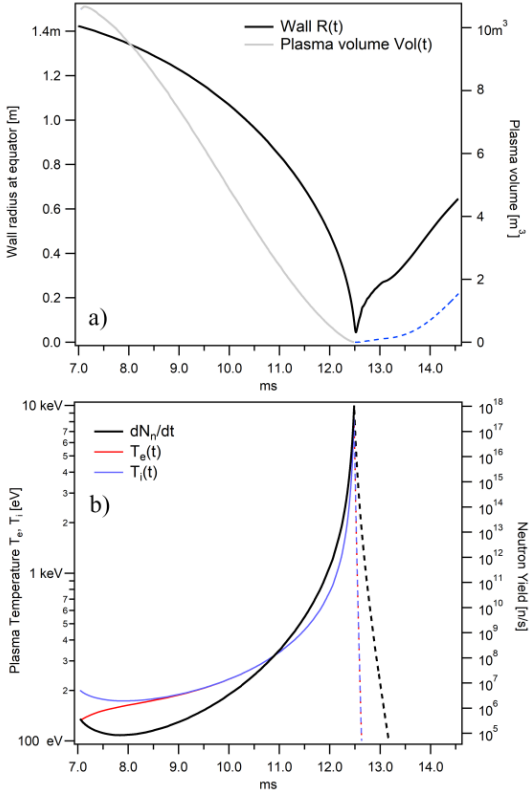

*Fig. 3. Example of compression scenario for FDP, a) Wall motion begins at t =0, plasma is injected at t = 7 ms, peak compression occurs at t = 12.5 ms. Exact timing values depend on details of compression parameters. b) Dynamic neutron yield increases by 13 orders of magnitude during compression as temperature and density rapidly increase. Dashed curves after peak compression represent a conservative estimate of decreased fusion due to cooling during rebound. FWHM of the neutron pulse is 12 μs. Direct neutron travel time from source to detector is 300 ns.*

## III. MCTOF DESIGN

The neutron spectrometer estimates $T_i$ from the distribution of $N$ neutron energy measurements. The relative uncertainty in $T_i$ is given by

$$\frac{\delta_{T_i}}{T_i} = \frac{2(\sigma_{E_n}^2 + \sigma_{T_i}^2)}{\sigma_{T_i}^2 \sqrt{2N-2}} \quad (2)$$

where $\sigma_{T_i}$ measures the broadening of the distribution due to the ion temperature and $\sigma_{E_n}$ is the standard deviation of the distribution due to the system's finite energy resolution (measured σ for $T_i=0$ case).

A neutron scatters elastically at an angle $\theta$ from Layer 1 to Layer 2 and travels a radial distance $R = \sqrt{X^2 + Y^2}$, and an axial distance Z, in a time $t_{TOF}$. The energy of the incident neutron is

$$E_n = \frac{m_n(R^2 + Z^2)^2}{2Z^2 t_{TOF}^2} \quad (3)$$

where $m_n$ is the neutron mass. The energy resolution of the system can be estimated from propagation of uncertainty.

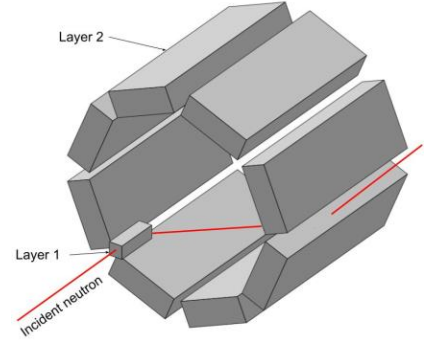

*Fig. 4. Multilayer Coaxial Time-Of-Flight (MCTOF) conceptual 3D model. A neutron trajectory which scattered in both scintillator layers is shown in red.*

A measurement of $T_i$ is needed at peak compression, when the expected peak D-D yield is $10^{18}$ neutrons/s, and the TOF energy spectrum will be accumulated from neutrons passing along the unobstructed line of sight through the collimated hole in the shaft as shown in Fig. 5. The distance from Layer 1 scintillators to the plasma is 6 meters, and the hole through the shaft has a diameter of 2 cm. This gives an instantaneous neutron flux of $10^9$ neutron/s at Layer 1. Neutron and gamma shielding directly surround the MCTOF spectrometer.

During the 10 μs at peak compression there will be ~10,000 D-D fusion neutrons that will pass into Layer 1 at the exit of the collimator. The overall efficiency of the system is estimated to be around 5%, giving a count of 500 TOF coincidence events out of which to compose an energy spectrum and an average count rate of 50 MHz. One clear challenge with such a high neutron count rate is the increased chance of neutron pulse "pile-up" as well as an issue of ambiguity in matching events in Layer 1 to events in Layer 2. The pile-up problem can be addressed by segmenting the scintillators so there is a lower chance of multiple neutron events producing overlapping signals in the same detector channel.

For event ambiguity, while it is possible to exclude some events in post-data analysis according to pulse height [12], an intrinsic improvement can be found by correctly sizing the overall TOF distance between the layers to match the expected peak count rate. To avoid the ambiguity of multiple coincidence pairs overlapping in time with each other, there is advantage in minimizing the TOF between the two layers, so that the crossing time of one neutron is complete before the next neutron is likely to arrive at the spectrometer. However, a shorter overall TOF increases the relative uncertainty in each $t_{TOF}$ measurement, which will increase the uncertainty in the energy resolution unless it can be compensated for by decreasing the spatial uncertainty of scattering events. As shown in Fig. 6, the baseline MCTOF configuration was chosen based on optimizing overall accuracy taking into account the trade-off between these different kinematic effects (Figs. 7, 8).

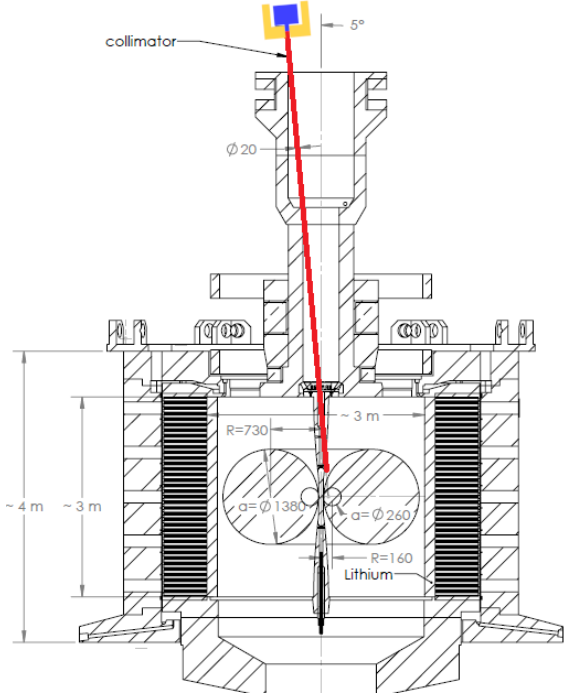

*Fig. 5. Proposed NES sight line for late compression at which point the lithium liner will obstruct nearly all other views of the plasma. MCTOF is the blue box, with shielding in orange.*

For Layer 1, Eljen EJ232-Q (0.5% Benzophenone) scintillator is chosen since it has a fast 0.11 ns rise and 0.7 ns fall time, which is important for the high neutron rates expected. The plastic scintillator is split so that there are 60 small bars each with dimensions 6.7 x 6.7 x 10 mm. These bars are layered 3 by 2 to span the neutron collimator and stacked 10 deep. Each Layer 1 bar has a single SiPM as the photo detector for registering neutron collisions.

Layer 2 has a large phase space for the neutrons to spread out, so Eljen EJ230 scintillator will be used with a slower 0.5 ns rise and 1.4 ns fall time. EJ230 also provides good light transmission and so is suitable for long, narrow bars. The scintillators will be grouped into 8 sections of 64 bars with an overall arrangement of an octagon. Each bar has dimensions 275 x 7.5 x 7.5 mm. The bars are packed in a 5 by 13 rectangle with each section missing one bar to keep to the standard 64 channels supported by the receiving electronics.

There are two SiPMs per Layer 2 scintillator bar, one on each end of the scintillator. When a neutron scatters in a Layer 2 scintillator, the two SiPMs each see a light pulse with different time delays. The time difference between the two light pulses determines axial position Z within the bar of where the scattering event occurred.

The center of the corresponding Layer 1 event and the (R, Z) coordinates of the Layer 2 event are used to determine the incident neutron energy $E_n$ via Eq. (3).

The uncertainty in the X, Y position of a scattering event depends on the dimensions $D_X$, $D_Y$ of the scintillator on which it was detected: $\sigma_{X_{1,2},Y_{1,2}} = D_{X_{1,2},Y_{1,2}}/\sqrt{12}$. The uncertainty in Z is dominated by the long bars in Layer 2, where the position is obtained from the time difference between the two SiPMs. Testing has shown a resolution better than 10mm for bars of similar length to the ones proposed here, so we use a value of $\sigma_Z = 10$mm. [13]

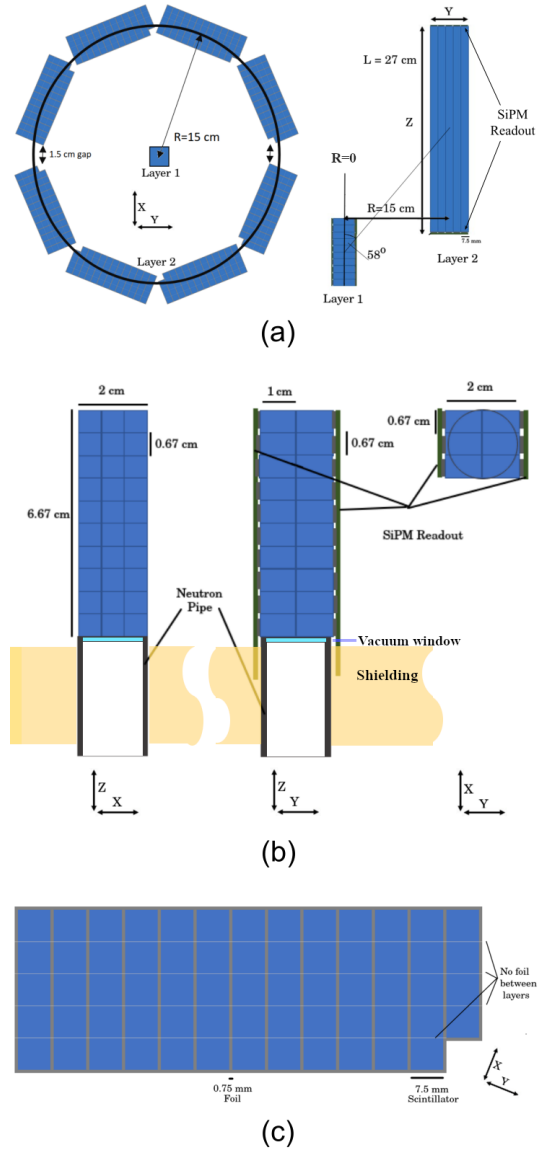

*Fig. 6. Baseline MCTOF concept for achieving high spatial resolution: (a) overall arrangement of detectors, (b) Layer 1 scintillator stack (subdivided into 60 small bars each with its own SiPM) on top of the exit pipe of neutron collimator, (c) Layer 2 scintillator block (one of 8), subdivided into 64 scintillator bars each with SiPM detectors on either end of bar for axial position localization from photon TOF within the 275 mm long bar.*

This design will require a total of 1084 SiPM detectors being read by a set of 17 ASIC boards (Liroc, 64 channels each, made by Weeroc [14]). The TOF uncertainty depends on the timing resolution of the electronics (~20 ps), the SiPMs (~50 ps) and the light travel time in Layer 2. A conservative estimate is taken to be $\sigma_{TOF} = 150$ ps.

Simulation with GEANT4 [15] was done to check analytical assumptions and evaluate the performance of MCTOF. The baseline geometry of the proposed scintillator configuration was used (Fig. 6). The angle between the centers of Layer 1 and 2 was varied to find an optimum. The optimum was found to be 58 degrees, which gives a relative temperature uncertainty of 13.6% with an efficiency of 6.2%. Simulation results were also compared with analytic estimates from the equations for energy and temperature resolution previously stated. The energy resolution as a function of scattering angle matched reasonably well between simulation and analytic calculation (Fig. 7).

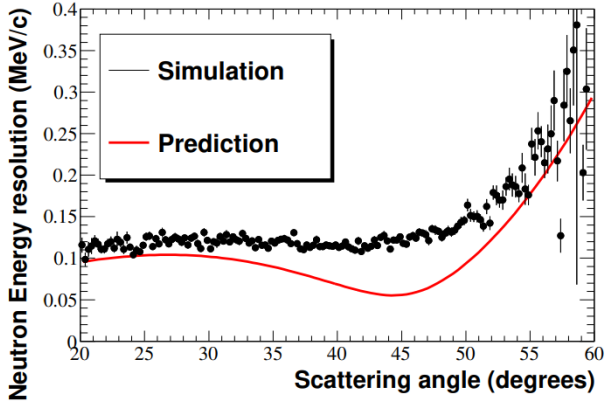

*Fig. 7. GEANT4 simulation results for the proposed MCTOF geometry with the baseline design parameters.*

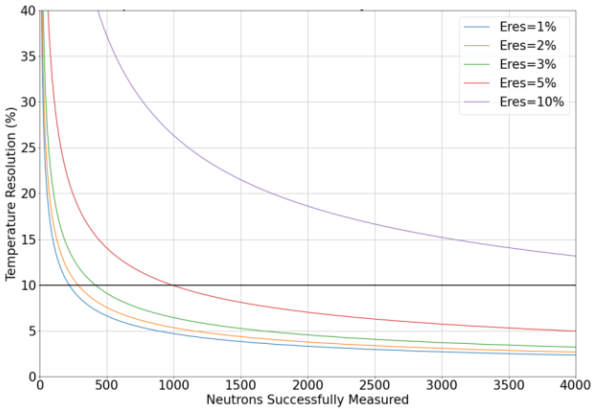

*Fig. 8. Analytic calculations of temperature resolution, $\delta_{T_i}/T_i$, for the baseline MCTOF configuration plotted as a function of N, the number of neutron coincidence pairs measured. A range of values for energy resolution, $\sigma_{E_n}/E_n$, are shown.*

The effects of background noise from gammas and back-scattered neutrons have not been fully analyzed. However, initial 3D MCNP simulations [16] with a simplified FDP geometry (Fig. 9) suggest the signal from gammas will be 2 to 3 orders of magnitude less than from neutrons, and back-scattered neutrons will make up much less than 10% of the signal from collimated neutrons in the energy range of interest. More work must be done to fully understand this potential source of noise.

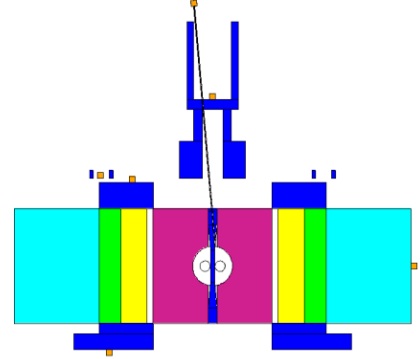

*Fig. 9. Simplified FDP geometry for 3D MCNP simulation of neutron fluxes at MCTOF. Pink: lithium. Dark blue: steel. Green, yellow: steel, with density appropriate to pistons at peak compression. Cyan: synthetic material with the average density of pistons and the rotor at peak compression. Orange boxes are scintillators.*

## IV. ACKNOWLEDGMENTS
Thanks to the Applied Nuclear Physics group at Uppsala University for insightful conversations on neutron emission spectroscopy.

## V. DATA AVAILABILITY STATEMENT
The data that support the findings of this study are available from the corresponding author upon reasonable request.

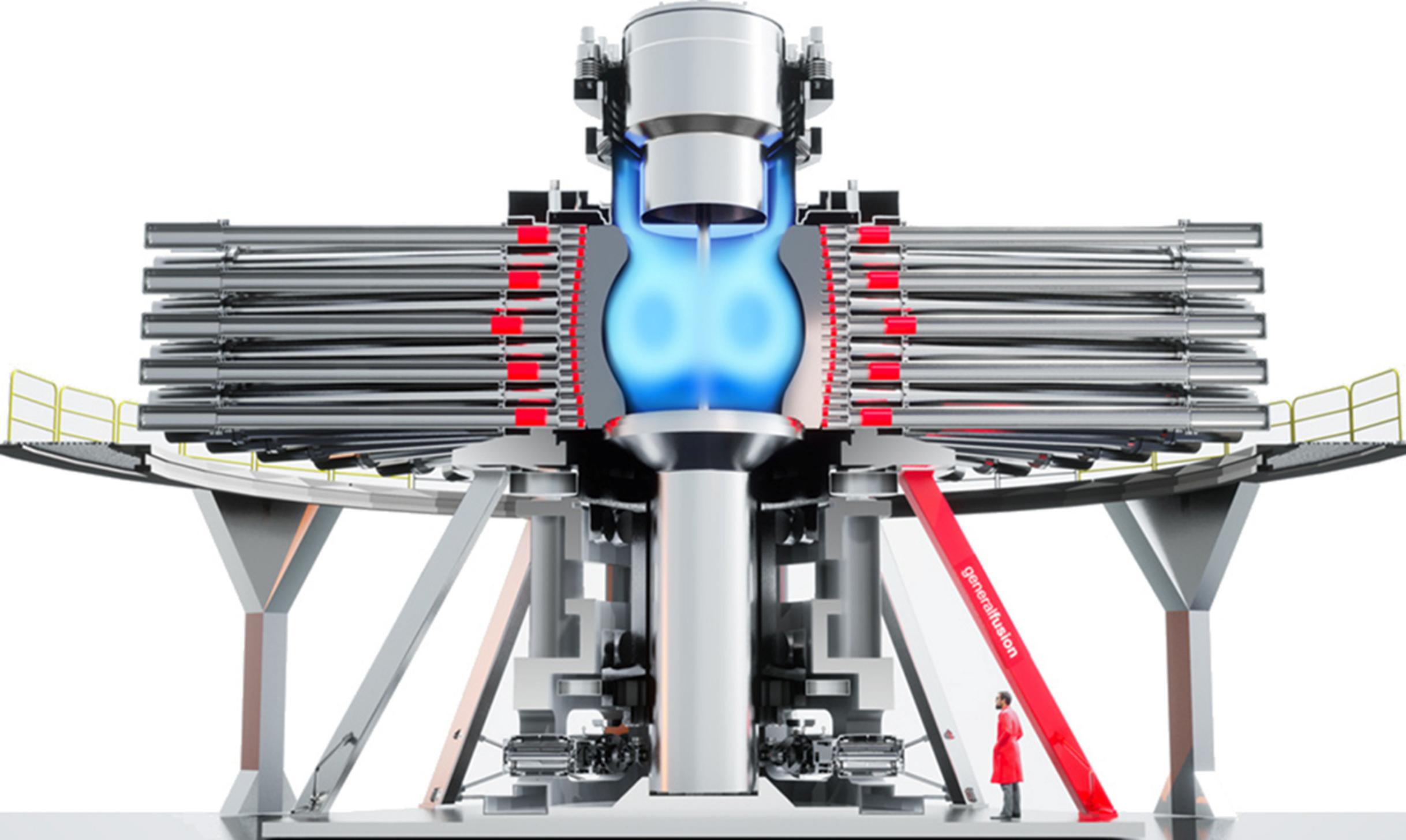

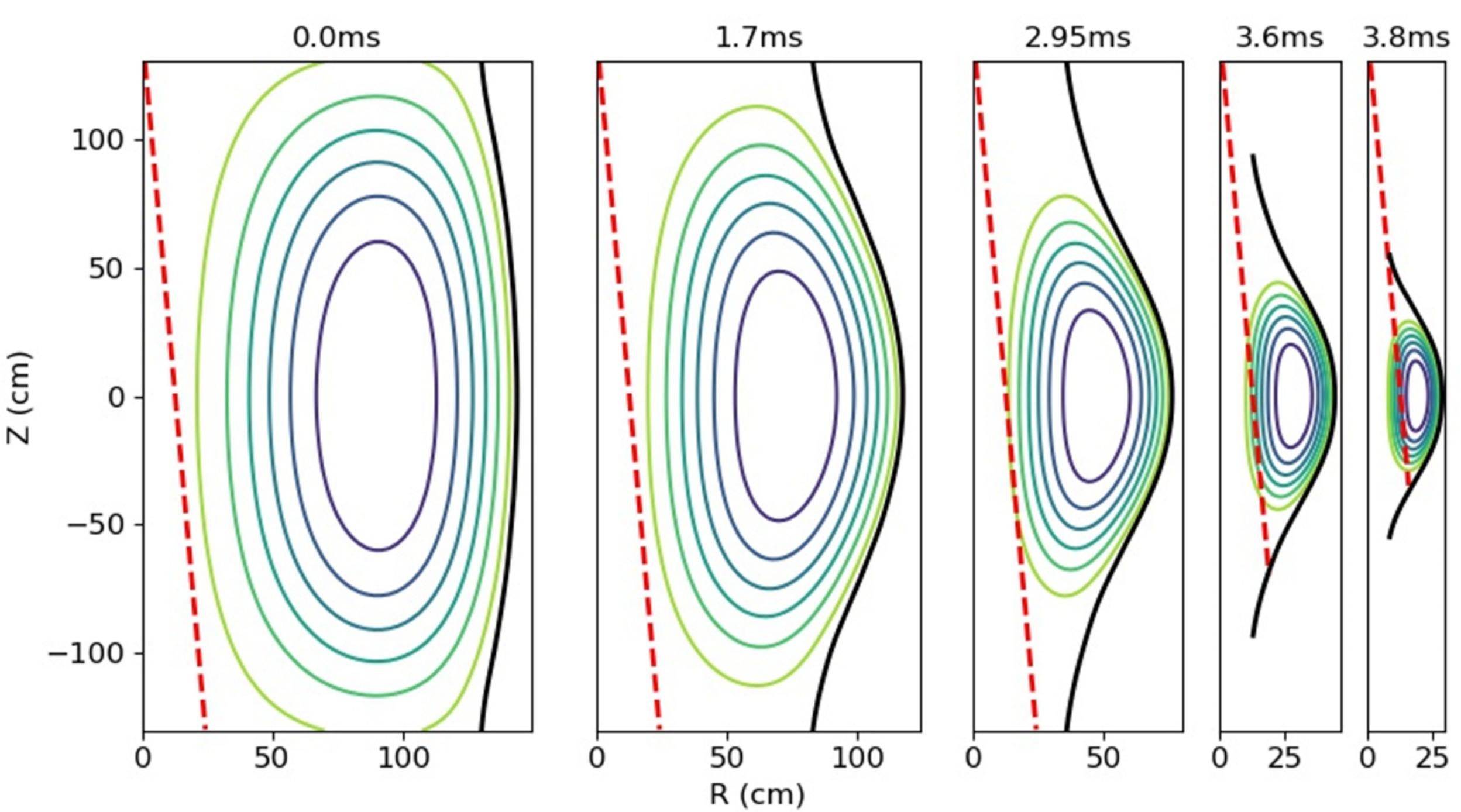

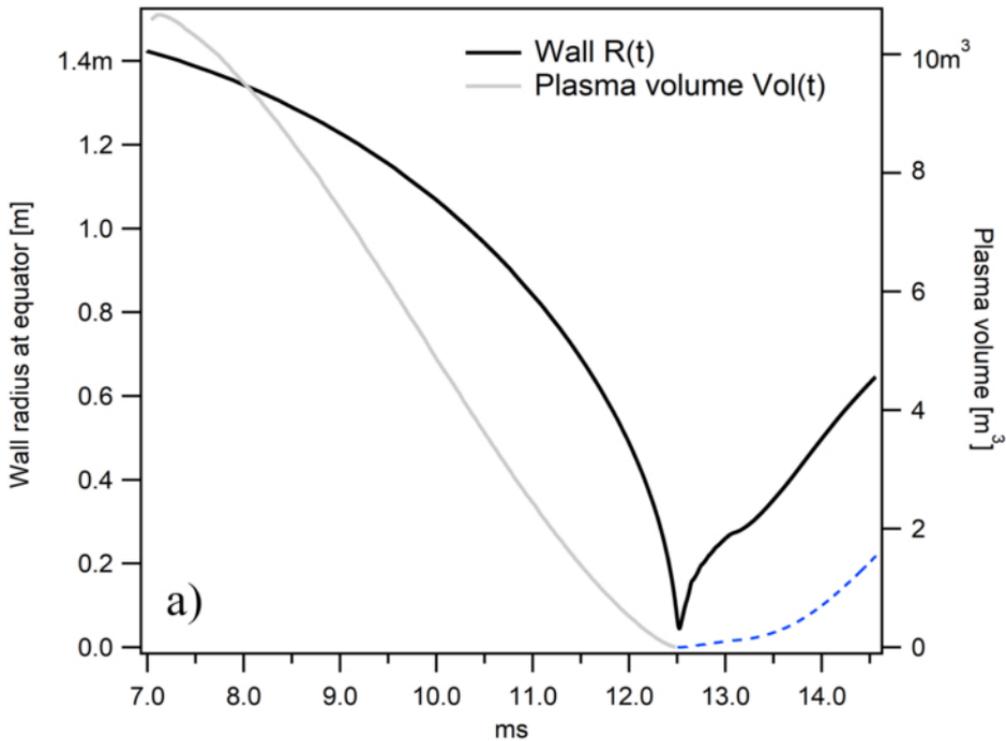
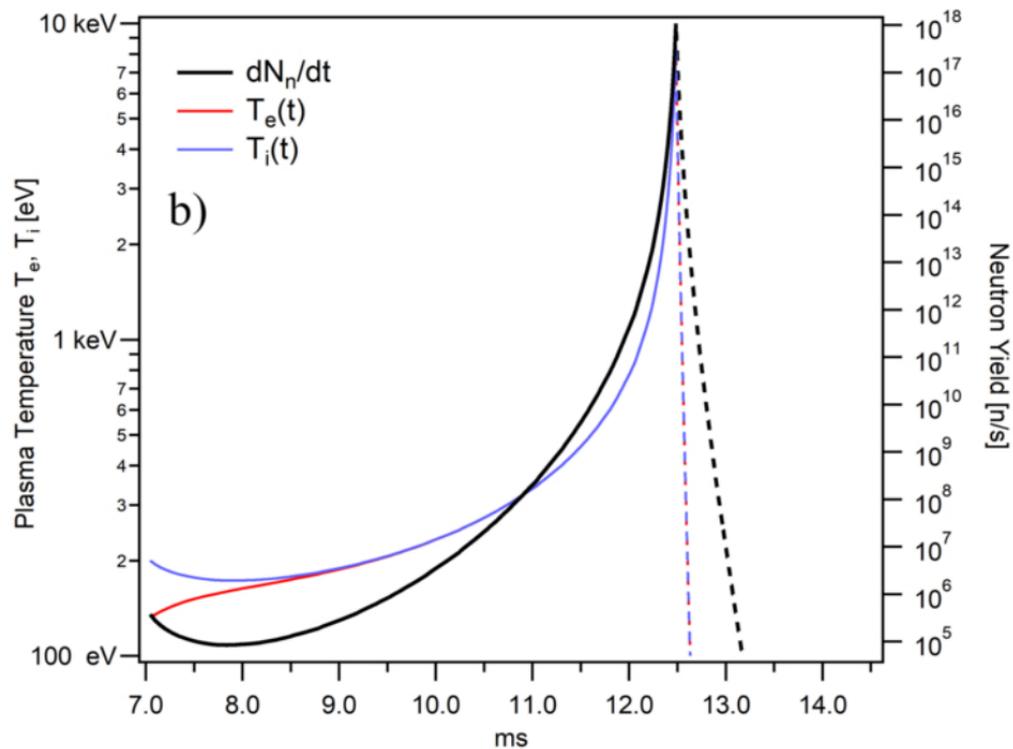

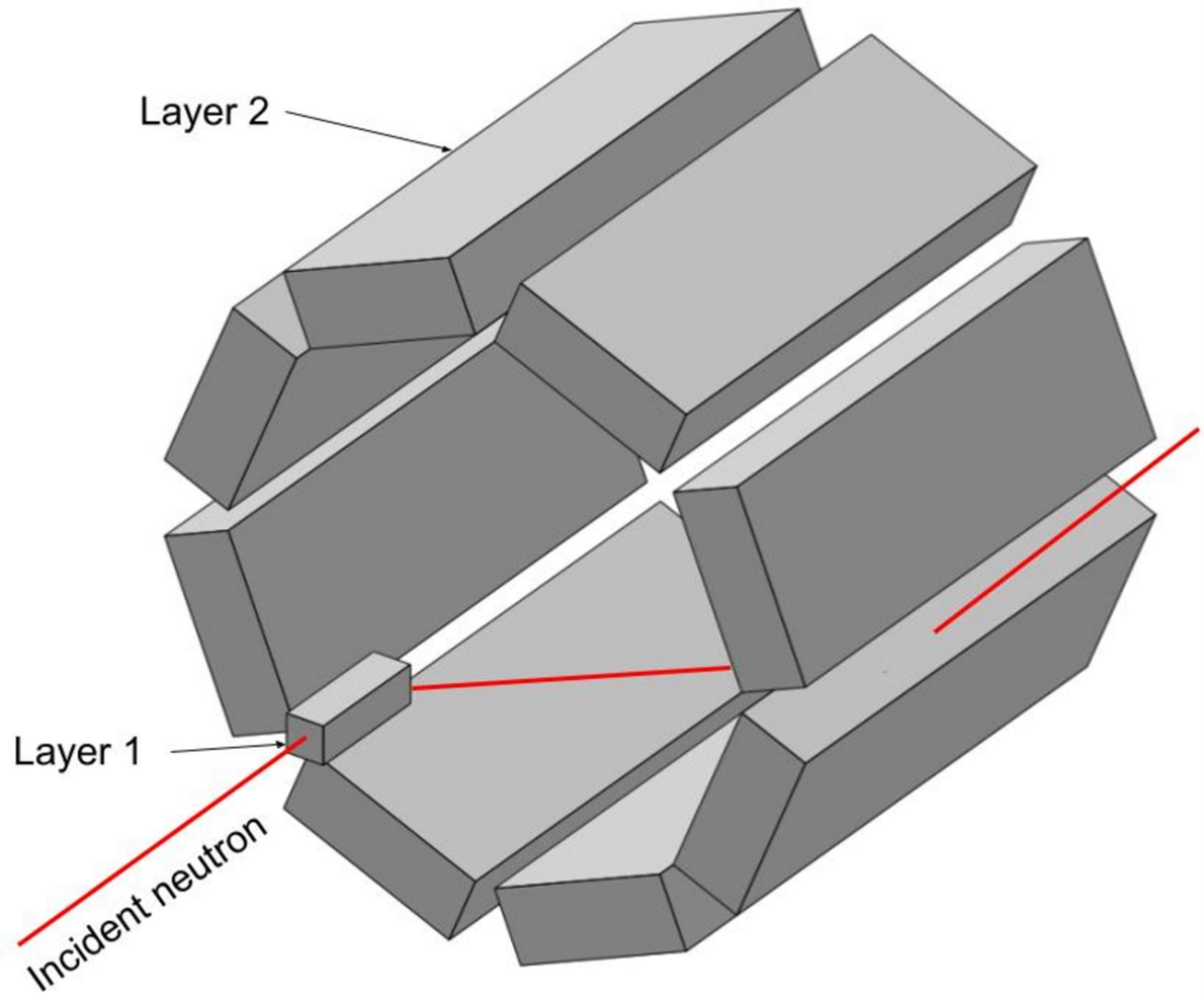

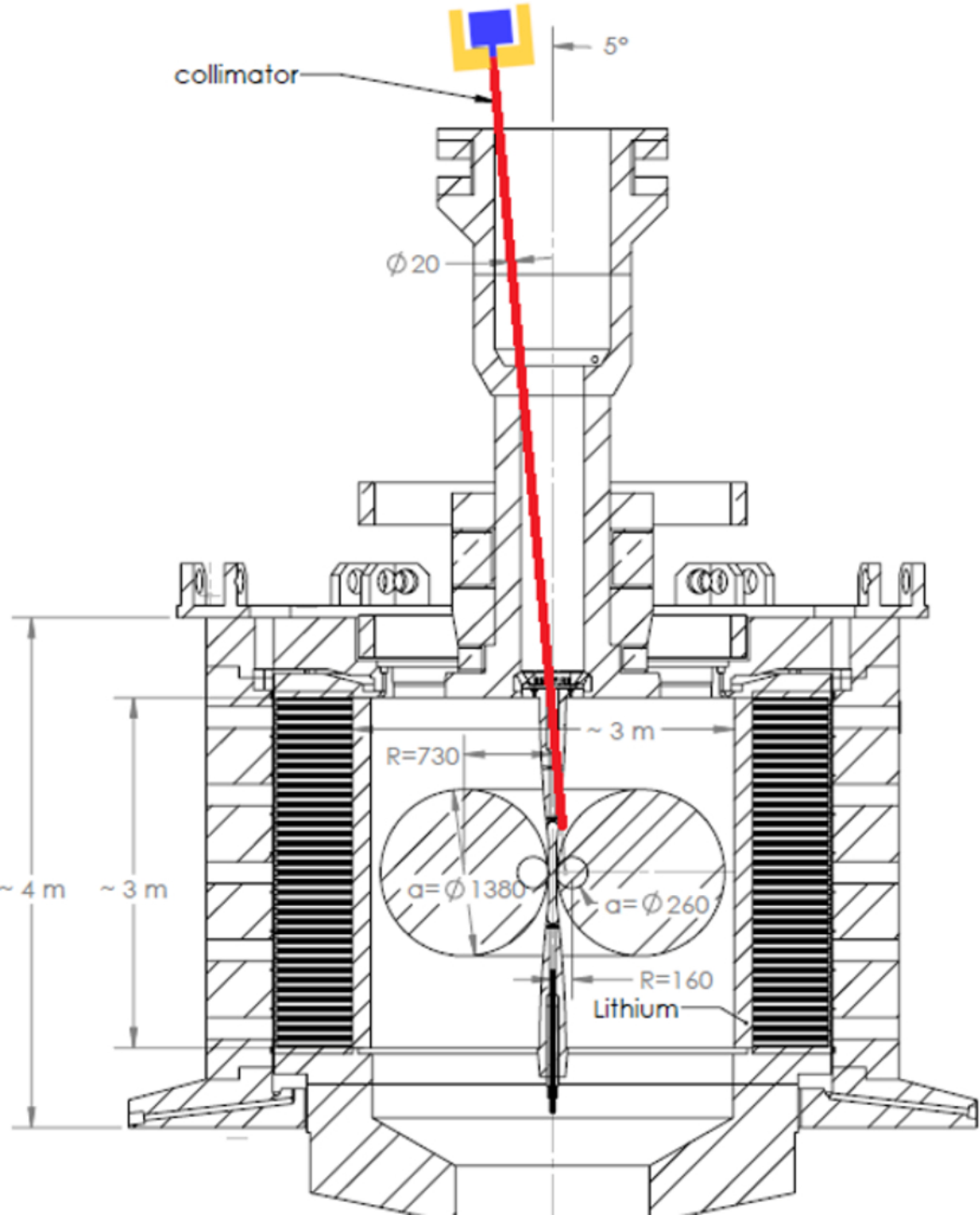

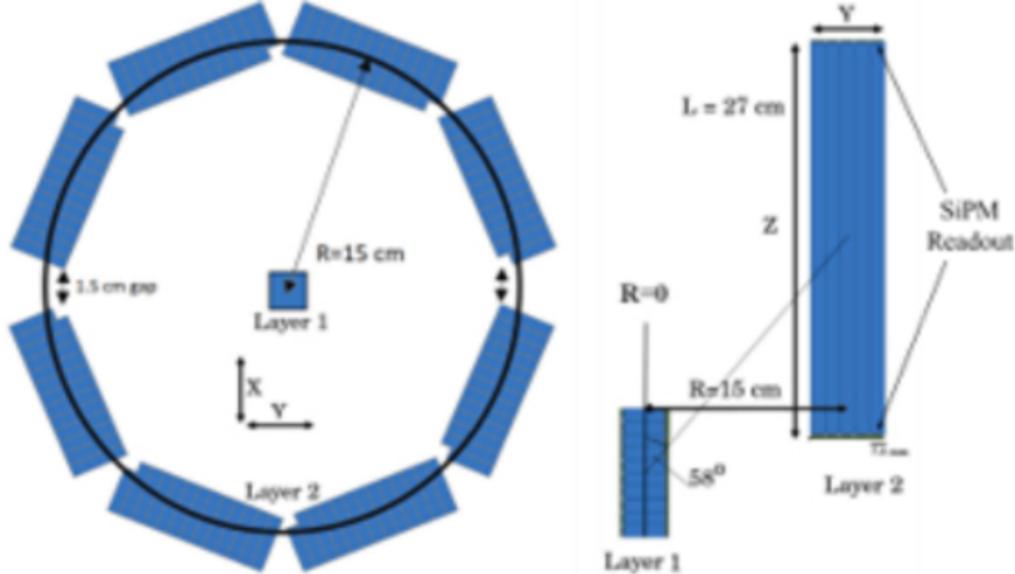

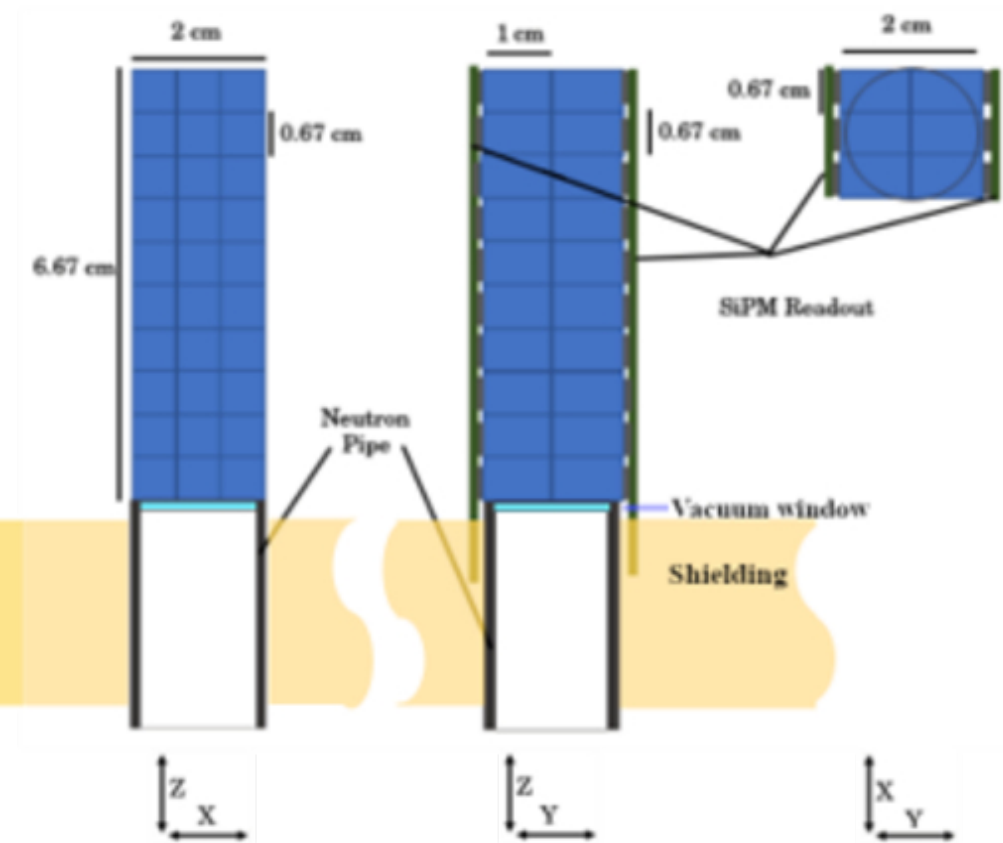

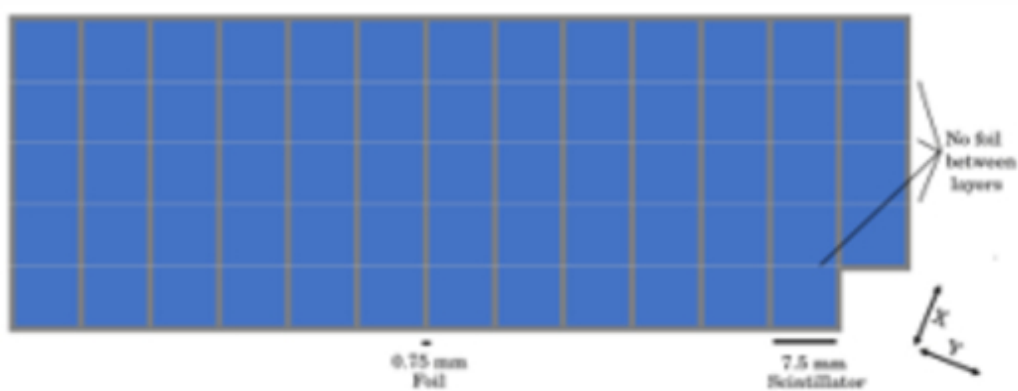

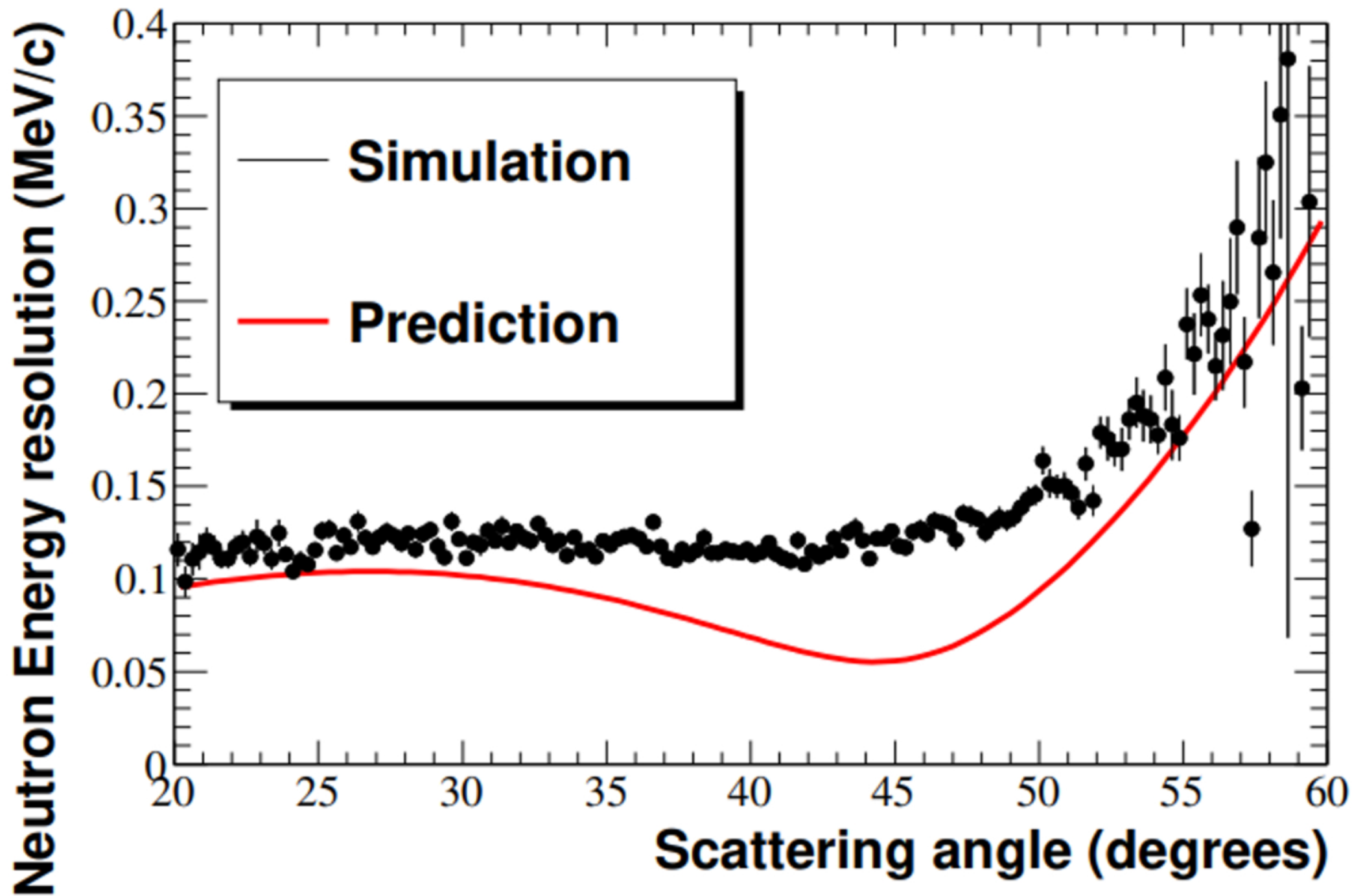

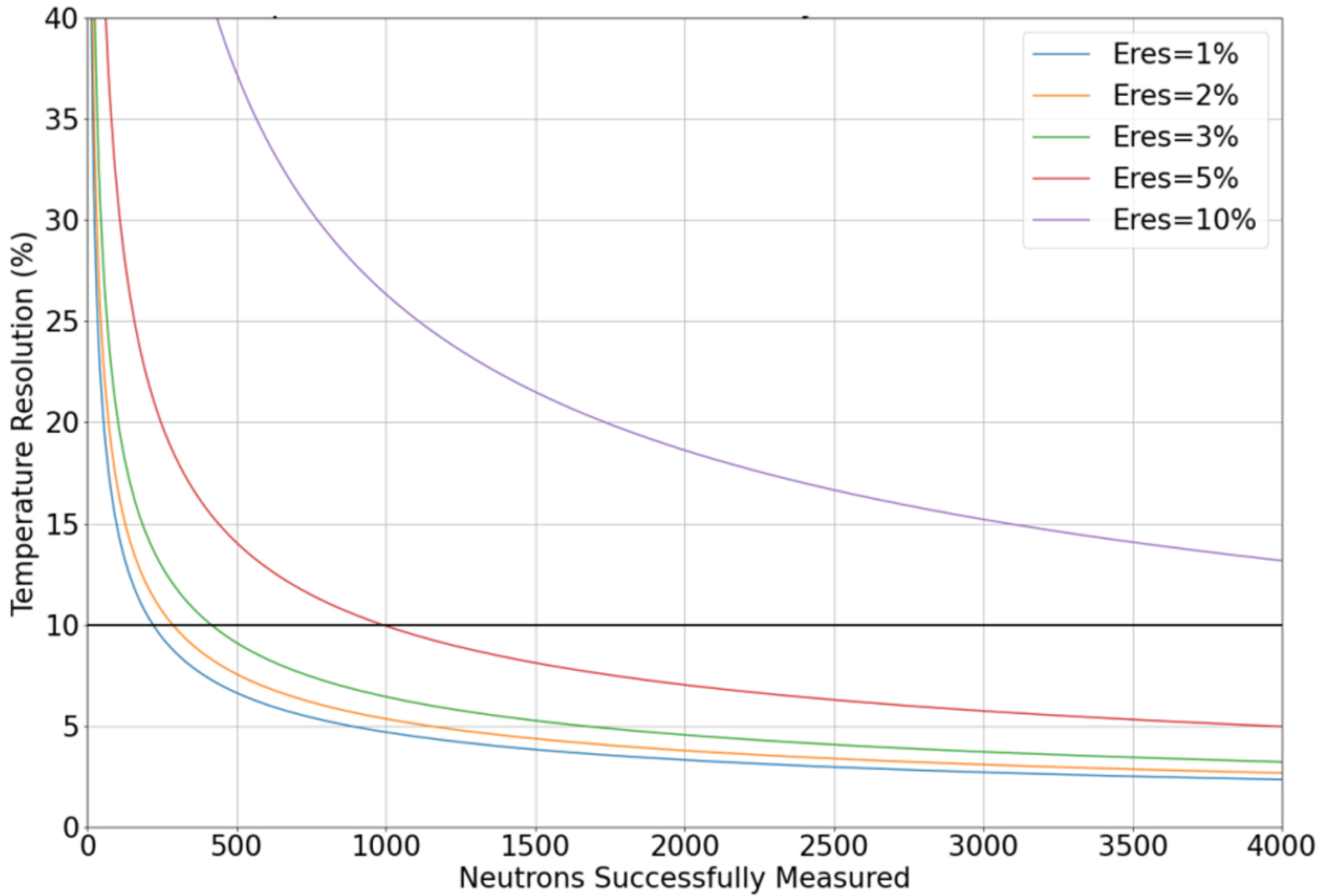

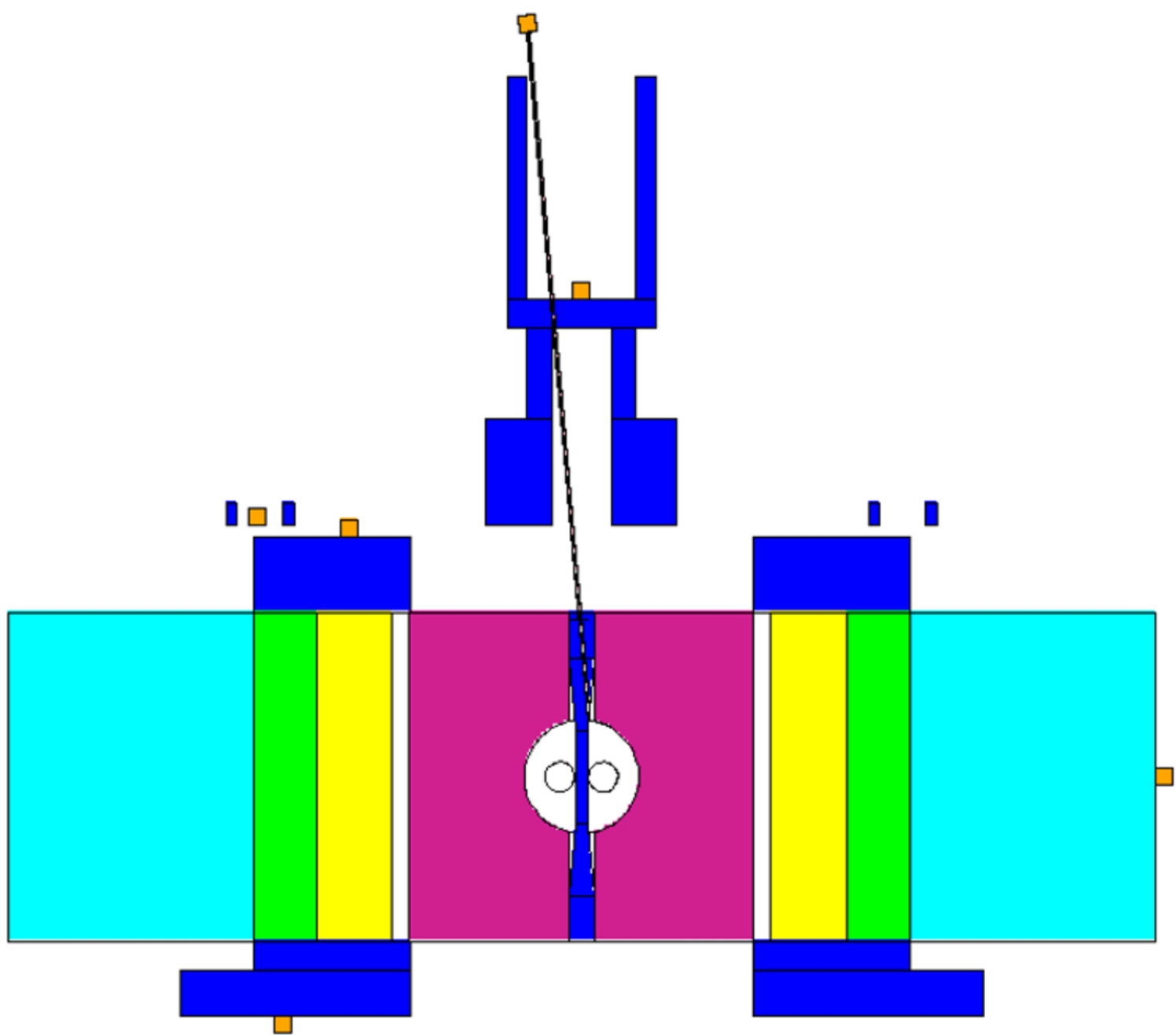